\documentclass[
reprint,
 amsmath,amssymb,
 aps,
pra,
]{revtex4-2}

\usepackage{graphicx}
\usepackage{tikz}
\usepackage{dcolumn}
\usepackage{bm}
\usepackage{braket}

\newcommand{\sugg}[1]{\textcolor{red}{#1}}
\let\sugg\relax

\begin{document}

\title{ \sugg{ Fundamental and operational limitations of Fisher-information-based quantum metrology }}

\author{Zden\v{e}k Hradil$^*$}
\affiliation{Department of Optics, Palack\' y University, 17.
listopadu 12,  779~00 Olomouc, Czech Republic}

\thanks{Corresponding author: hradil@optics.upol.cz }
\author{Jaroslav \v{R}eh\'{a}\v{c}ek}
\affiliation{Department of Optics, Palack\' y University, 17.
listopadu 12,  779~00 Olomouc, Czech Republic }


\date{\today}

\begin{abstract}
\sugg{ Quantum-enhanced sensing is commonly benchmarked using the quantum Fisher information (QFI), often interpreted as a direct indicator of achievable precision. However, such an interpretation is only justified within a fully specified inference framework that consistently incorporates state preparation, measurement design, resource accounting, estimator construction, prior information, and finite data effects. Here we develop an operational, end-to-end framework for quantum sensing under finite resources and identify general principles required for meaningful performance assessment. A central conceptual point is that the relevant unit of estimation is not a single detection event, but the complete data set required to construct a consistent estimator.   Rather than relying on abstract statistical formalism, we analyze a set of paradigmatic and experimentally relevant examples that expose common pitfalls in widely discussed quantum sensing strategies. These include quantum interferometry, sensing with squeezed states, and criticality-inspired protocols.
Our results clarify when large Fisher information or nonclassical resources translate into genuine metrological advantage, and when they do not. They further provide a practical methodology for designing and evaluating quantum sensing protocols under realistic experimental constraints. 
}
\end{abstract}

\maketitle


\section{\label{Intro} Introduction }

Quantum sensing is one of the most rapidly developing areas of quantum technology and is often celebrated for its potential to deliver a practical quantum advantage. 
This new era was inaugurated by the experimental detection of squeezed states with variances below the shot-noise limit. 
However, the mathematical foundations of the field reach much deeper into the past.

Estimation theory itself predates quantum mechanics. It was established as a rigorous statistical discipline by R.~A.~Fisher more than a century ago~\cite{Fisher_22}, independently and contemporaneously with the birth of quantum theory. 
Several decades later, the formulation of the Cramér--Rao  (CR) inequalities by C.~Rao and H.~Cramér~\cite{Rao_45,Cramer_46} completed the core of the classical estimation framework that is still used today—often implicitly—in quantum metrology.

A major milestone in the foundations of quantum metrology and quantum sensing was achieved by C.~W.~Helstrom~\cite{Helstrom_76}, who optimized the Fisher information over all possible measurements, giving rise to the QFI, $F_Q \geq F$. 
These results initiated, at the beginning of the new millennium, an intensive search for probe states that maximize the attainable Fisher information. 
In this context, the so-called NOON states, introduced and studied by B.~Sanders \cite{Sanders1989}, J.~Dowling and P.~Kok \cite{Lee2002}, are commonly regarded as a paradigmatic benchmark~\cite{bouwmeester_2004,mitchell_2004}. 
In the present work, NOON states are used as an archetypal reference example, rather than as the sole target of our analysis.

Our central contribution is to demonstrate that several widely used performance guarantees—often illustrated by paradigmatic examples such as NOON states—rely on assumptions that are not operationally justified in realistic inference scenarios. 
Although such benchmarks are frequently invoked as indicators of quantum advantage, our analysis shows that the predicted enhancements need not be attainable, even in idealized and noise-free settings, once a consistent end-to-end inference problem is formulated. Importantly, our analysis does not question the validity of Fisher information as a statistical quantity. Rather, it clarifies the limitations of its common interpretation when used per detection event as a standalone indicator of achievable precision.

The  idealized benchmarks do not merely populate the theoretical literature. They increasingly shape experimental strategies, benchmarking practices, and the conceptual justification of contemporary quantum-sensing protocols across a broad range of platforms, including quantum optics~\cite{Nielsen_23}, quantum metrology~\cite{Demkowicz2012,Macieszczak_16,Tan_21}, optomechanics~\cite{Grochowski_25}, superconducting circuits~\cite{PRXQuantum}, solid-state qubits in NV centers~\cite{Yu2022QuantumFisher}, boundary time crystals~\cite{Cabot_24}, free-electron sources~\cite{Velasco_26}, and even cosmic qubits~\cite{ChenFeng2025_QFI_cosmic_qubit}. Across these diverse platforms, the QFI commonly serves as the ultimate quantum benchmark against which precision limits and quantum advantage are assessed. However, the operational attainability of this bound depends on nontrivial inferential assumptions that are rarely made explicit.

By making explicit the role of finite data, prior information, estimator construction, and resource accounting, our framework provides a corrective perspective directly relevant to both theory and experiment. 
\sugg{ In particular, we emphasize that shortcuts based on “ultimate limits” inferred from asymptotic limit of CR-type bounds do not necessarily correspond to physically achievable performance, as they may neglect essential aspects of resource constraints and finite data.
We illustrate these points through a set of physically motivated examples which, to the best of our knowledge, cover a broad class of sensing strategies discussed in the literature. }

\sugg{ We admit that there are alternative approaches capable of accounting for finite data sets based on Bayesian inference \cite{Rubio2020,Meyer26}, where uncertainty is naturally characterized in terms of posterior probabilities conditioned on explicit data sets. The purpose of the present Article is not to analyze conceptual differences between Bayesian and frequentist methodologies; rather, we view them as complementary approaches addressing partly different inferential questions.}

\sugg{ An estimator provides a point prediction—answering the question “what is the result?”—and plays a natural and important role, particularly in the early, discovery-oriented phase of an experiment, where identifying a plausible value of the parameter is often the primary objective. Point estimators are often attractive because of their operational simplicity and, in asymptotic regimes, may admit compact descriptions through moments and variance-based measures. The role of the estimator is indispensable, particularly in multiparameter settings with constraints imposed by quantum mechanics.}

\sugg{ However, interval-based inference addresses a complementary and equally important question: “which values remain compatible with the data?” In many areas of physics and engineering, this second question is often implicitly reduced to error bars associated with point estimates. In particular, we highlight the role of interval estimation—confidence, tolerance, prediction, or Bayesian credible intervals—as a natural and practically relevant framework for uncertainty quantification. Bayesian formulations themselves naturally lead to posterior regions conditioned on concrete data sets, reinforcing the importance of uncertainty descriptions that go beyond isolated point estimates or low-order moments. }

\sugg{ In technologically mature fields operating at the limits of performance, such as aerospace engineering or extreme ultraviolet lithography, uncertainty is routinely expressed in terms of such interval-based descriptions, as they directly inform decision-making under finite data and complex system dynamics. These descriptions are essential for reliability, certification, and control in high-stakes environments. Bridging the gap between point and interval estimation is therefore not merely a matter of statistical preference, but a prerequisite for the transition of quantum sensing from proof-of-principle demonstrations to deployable technologies. }

\sugg{ Finally, we briefly comment on the interplay between Bayesian and frequentist viewpoints. Rather than advocating a strict preference, we adopt a pragmatic perspective exploiting the complementary strengths of both approaches. As illustrated by our examples, the distinction is not merely methodological but reflects different inferential emphases: local parameter sensitivity, naturally associated with point estimation and Fisher information, versus global identifiability and compatibility with finite data, naturally addressed through Bayesian formulations and interval-based descriptions. } 


Our goal is not to scrutinize individual published results based on Fisher information, but rather to analyze representative examples in order to establish a common and operational benchmark framework that can be applied consistently across different sensing paradigms. As a central theme, we emphasize the role of well-defined benchmarks, resource accounting, prior information, and, crucially, the explicit construction of estimators---not merely ultimate precision bounds.
In this context, our analysis complements a small number of earlier critical studies demonstrating that claims such as so-called sub- or super-Heisenberg scaling cannot, in general, be justified without additional assumptions~\cite{Giovanetti_12,Peze_13,Meyer26,Criticality}. 
In particular, we show that the Heisenberg scaling commonly attributed to NOON states relies on strong prior information, and that a high value of the quantum Fisher information does not necessarily imply any operational quantum advantage once finite resources and a fully specified inference task are taken into account.

The article is organized as follows. 
In Sec.~\ref{Fisher}, we briefly review the concepts of Fisher information, quantum Fisher information, and the Cramér--Rao bound. We also introduce a consistent notation for resource accounting required to achieve a given resolution, and explain why considerations based on single-shot detection are not operationally meaningful. 
In the subsequent sections, we analyze representative quantum sensing protocols, including quantum interferometry (Sec.~\ref{Interferometry}), sensing with squeezed light (Sec.~\ref{Squeezed}), and the role of criticality (Sec.~\ref{Criticality}). These examples span a broad spectrum of quantum sensing scenarios and demonstrate that a complete estimation strategy is essential for a meaningful assessment of potential quantum advantages. Sec. \ref{Sum} highlights  the main methodological points for using Fisher information as meaningful operational measure under realistic conditions.
Conclusions are presented in Sec.~\ref{Con}.

\section{\label{Fisher} Fisher information, resources, and yield}

We do not aim to revisit the full mathematical formalism. Instead, by examining
formulations and arguments routinely used in the current literature, we highlight
conceptual shortcuts and practical limitations of commonly employed procedures.

Modern quantum sensing literature often relies on a highly simplified
interpretation of statistical estimation theory. In particular, the QFI \emph{per detection event} is frequently taken as the primary
performance metric, even though its operational meaning depends crucially on
assumptions that are rarely satisfied in realistic experiments.

The central result is the CR inequality, which relates the variance
of any unbiased estimator $(\Delta \theta)^2$ to the Fisher information $I_n$:
\begin{align}
    (\Delta \theta)^2 \ge \frac{1}{I_n(\theta)} = \frac{1}{n F(\theta)},
    \label{CR}
\end{align}
where $n$ is the number of detected events used to construct the estimator, and
$F$ denotes the Fisher information per detection event.

It is important to emphasize a frequently overlooked conceptual point. In the
derivation of the CR bound, the likelihood function is defined for the \emph{full
data set}, and the corresponding Fisher information refers to the entire sample.
The commonly used quantity $F$ arises only through the decomposition
$I_n(\theta)=nF(\theta)$. While formally correct, this decomposition obscures the
fact that the theory itself does not specify how large the data set must be for
the asymptotic regime—under which the CR bound becomes operationally
meaningful—to be reached. Focusing solely on $F$ therefore hides the crucial role
played by the sample size required to construct a reliable estimator.

We consider a quantum state $|\psi\rangle$ encoding an unknown parameter
$\theta$ via a unitary transformation
\[
U = e^{i \theta G},
\]
where $G$ is the generator. The transformed state is measured using a
positive-operator-valued measure (POVM) $\{\Pi_k\}$, yielding outcome
probabilities
\begin{align}
    p_k(\theta) = \langle \psi(\theta)|\Pi_k| \psi(\theta) \rangle .
\end{align}
The corresponding Fisher information $F$ (for multinomial statistics)  and quantum Fisher information $F_Q$
per detection event are given by
\begin{align}
    F &= \sum_k \frac{\big(p_k'(\theta)\big)^2}{p_k(\theta)}, 
    \qquad
    F_Q = 4 (\Delta G)^2 .
    \label{QFI}
\end{align}

While these expressions are standard, their interpretation requires care. Any
theoretical analysis should be complemented by the explicit construction of an
estimator $\hat{\theta}$ and verification of its scaling relative to the CR bound.
A consistent accounting of resources is essential in this context.

Let $r$ denote the resources required for a single detection event. Repeating the
experiment $n$ times consumes total resources
\[
N = n r .
\]
The CR bound can then be expressed as
\[
(\Delta \theta)^2 \ge \frac{1}{n F} = \frac{1}{N\, (F/r)}.
\]
In the quantum case, replacing $F$ by $F_Q$ yields an apparent enhancement.
However, this improvement becomes operational only if the scaling of $F_Q$ with
$r$ compensates the statistical cost of constructing the estimator.

In particular, Heisenberg-like scaling $(\Delta \theta)^2 \sim 1/N^2$ requires
\[
F_Q \sim r^2,
\]
together with a sufficiently large number of samples $n$ to reach the asymptotic
regime. This condition is highly nontrivial, as the required sample size itself
depends on the structure of the underlying probability distribution.
The origin of this constraint is intuitive. Large Fisher information is typically
associated with rapidly oscillating likelihood functions in the parameter
$\theta$. However, constructing an estimator with small variance requires a
sufficient number of samples to resolve these oscillations and to reach the
Gaussian regime in which the Fisher information governs the variance. These
requirements are inherently in tension.

Importantly, the number of samples $n$ required for constructing a well-behaved estimator
cannot be inferred directly from the Fisher information per detection. This issue
is well known in mathematical statistics in the context of finite-sample
performance beyond the CR bound. However, practical resolutions typically
rely on model-specific assumptions that are difficult to justify in realistic
physical settings. Arguments based on the sampling theorem and noise analysis
are discussed in Sec.~\ref{Criticality}.

A conceptually consistent experimental strategy is therefore to avoid relying
solely on asymptotic bounds, and instead to explicitly construct estimators and
evaluate their performance for finite data sets. Statements such as
“...we focus on maximizing the asymptotic single-shot Fisher information...”~\cite{Grochowski_25}
risk misinterpreting the role of the estimator. An undersampled estimator may
perform poorly, while an oversampled strategy  eliminates any apparent quantum
advantage by effectively reverting to a classical regime.

\sugg{
In this sense, the natural unit of estimation is not a single detection event but the full data set required to construct a reliable estimator. Without explicit estimator construction and proper resource accounting, claims of Heisenberg-like scaling cannot be operationally substantiated and may lead to misleading conclusions. Ongoing efforts to refine Heisenberg limits through corrected resource-counting frameworks further underscore that the identification of physically meaningful resources remains a subtle and still actively debated aspect of quantum metrology \cite{Gorecki2020}. }


A more experimentally meaningful quantity is the variance of an estimator constructed from all data collected within a given time window, or, more generally, any cumulative observable such as an integrated dose~\cite{dose}.
Here, beyond the required number of detections, the \emph{signal yield} $Y_{\psi}$ plays 
a crucial role.  
Exotic states—e.g., fragile superpositions—may be generated only with a low success probability 
$p_{\mathrm{gen}}$.  
In such cases, even if the QFI per detection is large, as in the example of coherent 
effects~\cite{Hradil_19}, where the vanishing signal strength fundamentally limits resolution.  
The overall yield of state preparation should incorporate further experimental aspects 
of state engineering, such as coupling (propagation) efficiency $\eta_{\mathrm{coup}}$ 
and detection efficiency $\eta_{\mathrm{det}}$, forming
\begin{align}
    Y_{\psi} = p_{\mathrm{gen}}\, \eta_{\mathrm{coup}}\, \eta_{\mathrm{det}}. 
\end{align}
Ultimately, what matters in practice is the effective Fisher information rate per time 
modified by the  yield  as $ n Y_{\psi} F . $  Non-Gaussianity is a clear obstacle due to the low yield and large cost  to build an estimator.

{ Fisher information is not a final result, but merely an intermediate tool for the design of an experimental setup. In the next section, we demonstrate how prior information and data set needed for building an estimator fundamentally limits the achievable resolution—an effect that lies completely beyond the reach of Fisher-information-based analysis. }


\section{ Quantum interferometry with NOON states}
\label{Interferometry}

\begin{figure}[ht]
  \includegraphics[width=\columnwidth] {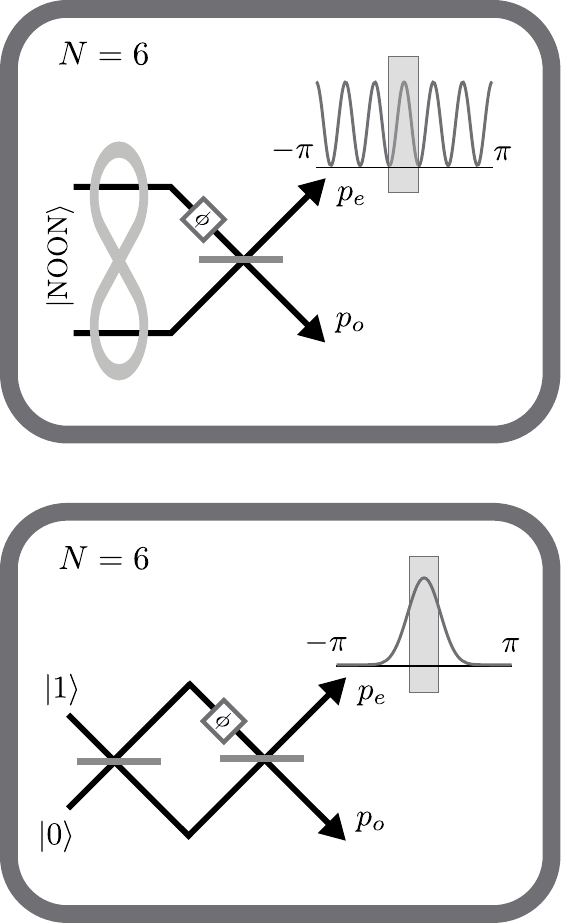} 
\caption{
Schematic of an interferometric phase-estimation protocol using an $N$-photon NOON state (upper panel)
and an $N$-times repeated standard Mach--Zehnder interferometer with a single photon (lower panel).
For phase estimation around zero phase shift, the posterior distributions scale as
$P_{\rm NOON} \propto \cos^2\!\left(\tfrac{N\phi}{2}\right)$ and
$P_{\rm MZ} \propto \cos^{2N}\!\left(\tfrac{\phi}{2}\right)$, as illustrated in the insets.
The apparent ``Heisenberg-like'' scaling does not originate from the detection scheme but is imposed
by prior information.
}
\label{fig_NOON}
\end{figure}

 {We start the analysis considering the   NOON state  as a candidate for reaching the Heisenberg-like scaling}
\begin{align}
    |{\rm NOON}\rangle = \frac{1}{\sqrt{2}}\big( |N,0\rangle + |0,N\rangle \big),
\end{align}  
injected into the two input ports of a beam splitter, with one mode subjected to a phase shift 
 $  e^{i \phi a^{\dagger}a}. $ The scheme is shown in Fig. \ref{fig_NOON}.
The detected photon-number outcomes $(k,N-k)$
 follow the multinomial distribution  $$  P(k) = \left(\tfrac{1}{2}\right)^N \binom{N}{k}\, \big[1 + (-1)^{N-k}\cos(N\phi)\big]. $$  
 This full photon-counting statistics can be reduced, without loss of information, to a binary (binomial) statistics distinguishing parity (even, odd) of  detection events, with probabilities
  \begin{align}  p_{e} = \cos^2\!\left(\tfrac{N\phi}{2}\right), \quad 
    p_{o} = \sin^2\!\left(\tfrac{N\phi}{2}\right) 
    \end{align}    
  No information is lost in this reduction, since both the QFI  and the classical Fisher information are equal,
$ F_Q=F=N^2 $
    corresponding to the Heisenberg scaling.

The limitation of a Fisher-information–based analysis becomes apparent at the level of estimator construction. Repeating the measurement does not allow one to identify a unique parameter value, since all maxima of the likelihood function are equivalent. As a result, the measurement outcomes cannot discriminate between parameter intervals separated by 
$2\pi/N.$
This intrinsic ambiguity originates from the oscillatory structure of the probability distribution (aliasing) associated with NOON states and is, in fact, the price paid for maximizing the Fisher information.

The enhanced Fisher information reflects only the local sensitivity around each maximum, while the global likelihood remains multi-modal. Introducing a finite prior lifts this degeneracy by restricting the admissible parameter range. However, this comes at the cost of eliminating any genuine quantum advantage in the global estimation scenario considered here: once the prior information fixes the interval, an equivalent resolution can be achieved even by classical or random sampling strategies.


For comparison, let us consider a standard Mach--Zehnder interferometer with a single photon in one input port while the other port is left empty, and detection repeated $N$ times. The counting statistics is described by a binomial distribution with probabilities
\begin{align}
    p_e = \cos^2\!\left(\tfrac{\phi}{2}\right), \quad 
    p_o = \sin^2\!\left(\tfrac{\phi}{2}\right).
\end{align}  

For a generic true phase shift, the observed data fluctuate, which in turn affects the estimation of the phase parameter. Here we illustrate, in simple terms, how the achievable precision should be assessed using both point estimators and Bayesian reasoning. First, note that the result depends strongly on the value of the \emph{true} parameter.
The most concentrated signal occurs at zero phase shift, where one of the output ports becomes dark. In this case, the binomial detection statistics becomes deterministic and the variance of the counted data vanishes. Within a local, point-estimation picture, this would suggest maximal sensitivity. However, this does not imply that the phase has been determined with arbitrary precision. The same detection record may arise for other phase values, just as repeated coin tosses may yield identical outcomes purely by chance.

The relevant question is therefore what can be inferred globally from such data. This can be addressed naturally within a Bayesian framework assuming a flat prior for the phase. Since the observed data are deterministic in this scenario, the uncertainty is fully captured by the posterior distribution, which takes the form
\begin{align}
    P_{\rm NOON}(\phi) \propto \cos^2\!\left(\tfrac{N\phi}{2}\right), \quad  
    P_{\rm MZ}(\phi) \propto \cos^{2N}\!\left(\tfrac{\phi}{2}\right).
\end{align}  

A convenient measure of dispersion is given by
\begin{align}
    D^2 = 1 - \left|\langle e^{i \phi} \rangle \right|^2,
\end{align}
which, for narrow distributions, scales as $D^2 \propto (\Delta \phi)^2$. Evaluating this quantity yields the same leading-order behavior
\begin{align}
    D^2 = 1 - \mathrm{sinc}^2\!\left(\tfrac{\pi}{N}\right),
\end{align}
corresponding to an effective phase uncertainty equivalent to a uniform distribution over the interval $(-\pi/N,\, \pi/N)$, with variance
\begin{align}
   (\Delta \phi)^2 = \frac{\pi^2}{3N^2}.
\end{align}  

These results are consistent with the Cramér--Rao bound~\eqref{CR} when expressed in terms of total resources $N$ (with $n=1$ and $r=N$). Importantly, the apparent Heisenberg scaling does not arise from the measurement itself, but is largely inherited from the prior constraint on the parameter range, with only minor refinement provided by the detection statistics.
 
There is another important aspect in the interpretation of information gain, namely the comparison between prior and posterior information. For NOON-state detection, the acquired information is operationally negligible in the resource-normalized sense considered here: the posterior distribution differs only marginally from the prior, whether one considers the full  $2\pi$
 interval or the reduced  $2\pi/N$   periodic domain. In this sense, the measurement does not lead to a meaningful information update. {This picture is fully consistent with interpretation of NOON states as  states with the "de Broglie" wave length $\lambda/N.$ Indeed, the measurement corresponds  to "single shot" detection on the interval of periodicity $2\pi/N$ with very little information gain since  scaling  of the variance $ 1/N^2$ comes from the prior assumptions.  }

By contrast, classical Mach–Zehnder interferometry does not provide additional information when restricted to the 
 $2\pi/N$ interval, but it yields a substantial information gain over the full  $2\pi$
range, as evidenced by a pronounced reduction of the posterior uncertainty.

\begin{figure}[ht]
  \includegraphics[width=\columnwidth] {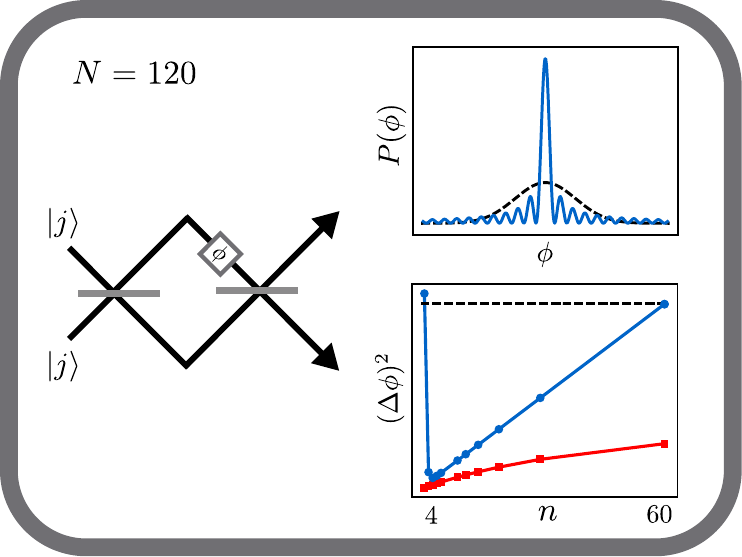} 
\caption{Schematic of an interferometric phase--estimation protocol based on the Holland--Burnett scheme. The phase is estimated at the operating point $\phi=0$ using $n$ repeated detections for  fixed  resources $N= 2n j .$ 
The upper inset compares the oscillating posterior phase distribution obtained from a single detection ($n=1$, solid blue line) with a conventional strategy in which all resources are injected into a single input port of a Mach--Zehnder interferometer (black dashed line). 
The lower inset  plots the variance of posterior distribution   as  a function of the number of repetitions $n$ (blue points). The optimal resolution is achieved for $n=4.$
The classical variance (black dashed line) and the hypothetical variance for  QFI (red points) are  shown for comparison. }
\label{HollandB}
\end{figure}

{ As the next example, we discuss a more realistic scheme often associated with Heisenberg scaling, namely the Holland--Burnett interferometer~\cite{Holland_93}, which can be viewed as a Mach--Zehnder interferometer fed by twin-Fock input states  and exploiting genuinely multi-photon quantum interference. In this setup (see Fig. \ref{HollandB}), two identical Fock states $\ket{j}$ are injected into the two input ports of the interferometer,  $r= 2j$ and the photon numbers at the output ports are measured as $ j \pm q. $   Denoting by $2q$ the detected photon-number difference, the conditional probability (likelihood) of observing the outcome $2q$ for a given phase shift $\theta$ can be well approximated for $q\ll j $ as
$p(q|\theta)\approx J_q^2(j\theta)$,
where $J_q$ is the Bessel function of the first kind. The optimal local performance is obtained when the (true) phase shift is set to $\phi=0$, for which the only   outcome is $q=0$.}

 Although the resulting phase-dependent likelihood is sharply peaked around $\phi=0$, it exhibits pronounced oscillations, which leads to a relatively broad Bayesian posterior distribution. In contrast to NOON-state statistics, however, the side lobes decrease with increasing $|\phi|$, so that repeated data acquisition prior to the final estimation step gradually suppresses the oscillatory structure and leads to a concentration of the posterior distribution. This effect is illustrated in the insets of Fig.~2. The optimal scaling can be inferred from the asymptotic behaviour of the Bessel functions, yielding an optimal repetition number $n=4$ for a total photon resource $N=2jn, r= 2j. $ For comparison Fisher and QFI  equal to $ F= F_Q= 2 j(j+1) $ implying the CR bound $(\Delta \phi)^2  \ge \frac{1}{N (j+1)}$~\cite{Hradil_05} which cannot be saturated but is well approximated by optimal regime.  
We conclude that this scheme is not as pathological as the NOON-state example, but it is still not compatible with a simple QFI-based diagnostic, since repeated sampling remains an essential part of achieving the optimal resolution and  repetition number $n$  is a physical optimization variable.

{ In realistic scenarios, however, the prior distribution cannot be concentrated at a single phase value but must be spread over a finite interval. This inevitably leads to fluctuations of the registered outcomes $q$ and, consequently, to a broader posterior phase distribution, which requires a larger data set in order to construct a reliable estimator. This analysis confirms that prior information about the estimated parameter and the size of the data set $n$ are irreducible components of any practically optimal estimation scheme. Diagnostics based solely on QFI  per detection cannot be justified.}

\section{Sensing with squeezed states}
\label{Squeezed}

\begin{figure}[ht]
  \includegraphics[width=\columnwidth] {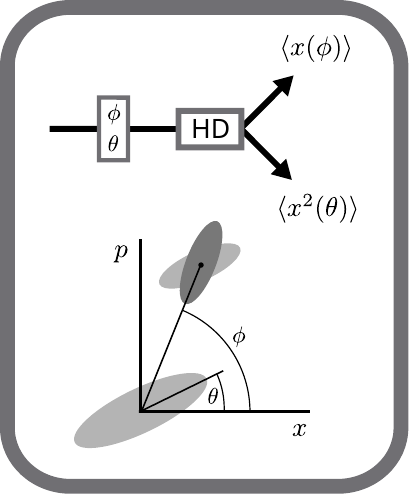} 
\caption{Schematic of homodyne detection scheme for bright and vacuum squeezed field. Notation of phase distinguishes between  phase of coherent amplitude $ \phi $ and phase of the squeezing $\theta.$   Whereas the the former one is  linked with Gaussian distribution, the latter one is related to  $ \chi^2 $ distribution. }
\label{HD}
\end{figure}

As the  example of a genuinely successful estimation strategy, let us consider the detection of the phase of a bright squeezed Gaussian state using homodyne detection, as sketched in Fig.~\ref{HD}.  
The problem of optimal phase resolution in this setting was analysed already in the early days of quantum optics and, to our best knowledge, the optimal scaling was first explicitly stated in~\cite{Bondurant_84}.

Since the statistics of the homodyne signal is Gaussian, a simple and transparent explanation can be given within linear error propagation, which is closely related to the arguments behind the Fisher information,
\begin{align}
(\Delta \phi)^2 \approx \frac{\Delta X(\phi)^2}{\bigl[X'(\phi)\bigr]^2},
\end{align}
where $X(\phi)$ denotes the rotated quadrature operator.  
The geometrical meaning of the optimal phase resolution is apparent from Fig.~\ref{HD}.  
If the anti-squeezed quadrature of a bright squeezed state is aligned with the complex amplitude, the phase uncertainty can be estimated as
\begin{align}
(\Delta \phi)^2 \approx \frac{e^{-2s}}{4|\alpha|^2},
\end{align}
where $\alpha$ is the coherent amplitude and $s$ is the squeezing parameter.

Using the relation for the mean photon number,
\begin{equation}
N = |\alpha|^2 + \sinh^2 s,
\end{equation}
the resolution can be adjusted to
\begin{equation}
(\Delta \phi)^2 \approx \frac{1}{4N^2},
\end{equation}
for the matching condition $|\alpha|^2 = e^{2s}/4$.  
Owing to the central limit theorem, the phase resolution obtained from repeated measurements depends only on the total amount of resources. Consequently, in this scenario the number of repetitions does not play an independent role if the total photon number is fixed. In this regime, the Fisher information provides a meaningful and operationally relevant figure of merit.

Two important points should, however, be emphasized.  
First, the optimal resolution is achieved only by combining classical and quantum resources: the squeezed state complements, rather than replaces, a bright coherent field.  
Second, attaining this performance requires the homodyne detection to be aligned with the mean amplitude, which represents a strong form of prior information about the phase.

An interesting question is whether a comparable strategy can be realized using purely quantum resources, such as squeezed vacuum. 
This problem was recently addressed in~\cite{Nielsen_23}, where a resolution ``beyond the NOON state'' was reported, 
with the analysis being based solely on the QFI per detection. 
Let us examine this scheme from a simple statistical perspective.

For a squeezed vacuum state, the QFI for phase estimation can be readily evaluated from Eq.~\ref{QFI} as
\begin{equation}
F_Q = 2\sinh^2(2s),
\end{equation}
but the question remains whether this quantity constitutes a statistically relevant measure of attainable precision.

Although a single quadrature measurement is Gaussian distributed,
$n$ repeated measurements do not lead to a Gaussian distribution for the quadratic estimator.
Instead, they give rise to a well-known $\chi^2$ distribution with $n$ degrees of freedom,
\begin{align}
p(y|\theta) &= \frac{1}{2^{n/2}\Gamma(n/2)}
\frac{1}{V(\theta)^{n/2}}
y^{\frac{n}{2}-1}
\exp\!\left[-\frac{y}{2V(\theta)}\right],\\ y &= \sum_{i=1}^n x_i^2, \quad
V(\theta) = V_{+}\cos^2\theta + V_{-}\sin^2\theta ,
\end{align}
and  $\theta$ denotes the true phase of the squeezed vacuum.
This scheme naturally arises in covariance-matrix estimation from homodyne data~\cite{Rehacek_2009},
since the random variable $y$ may serve as an estimator of the variance,
$V(\theta) \approx y/n$, and can, with appropriate care, also be employed for phase inference.

However, the induced statistical model is intrinsically nonlinear.
For sufficiently large $n$, the $\chi^2$ distribution admits an approximate Gaussian representation.
Using a Fisher-type square-root transformation, one obtains
\begin{align}
p(y|\theta) \propto
\frac{1}{V(\theta)^{1/2}}
\exp\!\left[
-\frac{\bigl[\sqrt{2y}-\sqrt{(2n-1)V(\theta)}\bigr]^2}{2V(\theta)}
\right].
\end{align}
This expression can be interpreted as estimation of the random variable
$\xi = \sqrt{V} \approx \sqrt{\frac{2y}{2n-1}}$ with variance
\[
(\Delta \xi)^2 = \frac{V}{2n-1}.
\]
Consequently, the relative uncertainty obeys
\[
\frac{(\Delta \xi)^2}{\xi^2} = \frac{1}{2n-1},
\]
which corresponds to purely classical scaling with the sample size.
The QFI scaling originates from the parameter dependence of the variance, whereas the estimator uncertainty obeys classical sampling scaling.

The situation becomes even more involved when realistic data-selection procedures
and estimator bias are taken into account in phase estimation.
Under such conditions, the Cramér--Rao bound~\eqref{CR}, derived under regular unbiased-estimator assumptions,
does not directly characterize the achievable precision.
A complete inferential analysis, including consistency conditions, admissible data domains,
and optimal estimator construction, lies beyond the scope of the present work and will be presented elsewhere.
Here we restrict ourselves to identifying the statistical mechanism responsible
for the breakdown of the QFI-based diagnosis, which illustrates the central point of this article,
namely that the (classical) resources required for constructing a reliable estimator
are essential and cannot be neglected.  This behavior is conceptually analogous to the NOON-state example discussed above:
a large quantum Fisher information does not by itself guarantee an operationally attainable precision
when the statistical structure of the estimator departs from the regular Gaussian regime
assumed in standard asymptotic theory.

\sugg{
\section{Criticality in quantum sensing} }
\label{Criticality}

\sugg{
A frequently invoked argument in quantum sensing is that strongly oscillatory
probability distributions lead to enhanced Fisher information and, consequently,
to improved estimation precision \cite{Criticality,Gietka_PhysRevLett.132.060801}. While this is formally correct at the level of
local sensitivity, we show on a simple analytical example that such behavior
places stringent demands on the data set required for constructing a reliable estimator.
}

\sugg{
Consider a family of probability densities of the form
\begin{equation}
p(x|a) = \frac{1}{Z}
\exp\!\left(-\frac{(x-a)^2}{2\sigma^2}\right)
\cos^2\!\left(\frac{N\pi(x-a)}{2\sigma}\right),
\label{eq:toy_model}
\end{equation}
where $x$ represents data, $a$ is the parameter to be estimated, $\sigma$ sets the width of the
envelope, and $N \gg 1$ controls the oscillatory structure. The normalization
constant $Z$ is independent of $a$ due to translational invariance.
}

\begin{figure}[ht]
  \includegraphics[width=\columnwidth]{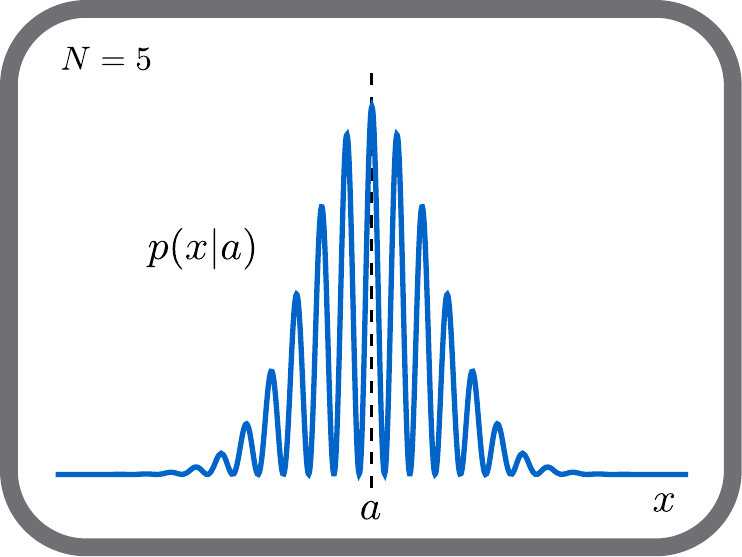}
 \caption{ \sugg{
  Schematic illustration of critical behavior arising from an oscillatory probability density $p(x|a)$. The corresponding Fisher information density,
$F(x|a)=\frac{(\partial_x p(x|a))^2}{p(x|a)}$, exhibits a complementary structure, with peaks occurring in regions where $p(x|a)$ is  vanishing. Consequently, the dominant contributions to the Fisher information originate from regions of low signal; see the main text.} }
  \label{Crit}
\end{figure}

\sugg{
The Fisher information associated with this model can be evaluated analytically
and scales as
\begin{equation}
\mathcal{F}(a) \sim \frac{1}{\sigma^2} + \mathcal{O}(N^2),
\end{equation}
demonstrating a rapid growth with $N$ due to the increasingly rapid oscillations
of the likelihood function. From a local perspective, this reflects strong
sensitivity to small parameter variations—a behavior often associated with
criticality (see Fig.~\ref{Crit}).
}

\sugg{
However, the same oscillatory structure induces a fundamentally different global
behavior. The likelihood function becomes highly multi-modal, with 
$N$ distinct maxima within the width $\sigma$ of the Gaussian envelope. These
maxima are separated by a characteristic scale
\begin{equation}
\Delta a \sim \frac{2\sigma}{N}.
\end{equation}
As a consequence, repeated detections cannot discriminate between parameter
values differing by integer multiples of $\Delta a$, and the data effectively sample only the Gaussian envelope.
}

\sugg{
The resulting estimator can therefore be interpreted as a mixture of locally
sharp distributions centered at  the cells $a + k\Delta a$, weighted by the Gaussian
envelope. Its variance can be decomposed as
\begin{equation}
\mathrm{Var}(\hat a)
=
\underbrace{\mathcal{O}(1/N^2)}_{\text{local sensitivity}}
+
\underbrace{\mathcal{O}(\sigma^2)}_{\text{global ambiguity}}.
\end{equation}
The first term reflects the large Fisher information, while the second term
originates from the inability to identify the correct cell. For $N \gg 1$,
the second contribution dominates, yielding
\begin{equation}
\mathrm{Var}(\hat a) \sim \sigma^2.
\end{equation}
}

\sugg{
From an operational perspective, resolving the ambiguity between different
cells requires additional resources. In particular, the statistical uncertainty
$\sigma/\sqrt{n}$ must become smaller than the spacing between adjacent maxima,
$\Delta a \sim 1/N$, which yields the necessary condition
\begin{equation}
n \gtrsim N^2.
\end{equation}
Only in this regime does the estimator enter the local Gaussian domain where
the Fisher information becomes operationally relevant.
It is important to emphasize that the frequently cited $1/N^2$ scaling should be
interpreted with care. In the present setting, this scaling reflects resolution
at the level of individual cells and becomes meaningful only after a sufficient
number of samples $n \sim N^2$ is acquired. Expressed in terms of total resources,
this corresponds to the standard $1/n$ scaling.
}

\sugg{
The crucial question in ``quantum sensing enhanced by criticality'' is therefore
whether the data set is sufficient to resolve the oscillatory structure of the
likelihood. If this condition is satisfied, the estimator may enter a regime in
which the $\mathcal{O}(N^2)$ contribution to the Fisher information leads to
improved precision. If not, the oscillatory structure is effectively averaged
out, and the estimation error follows the classical scaling determined by the
data sample.
}

\sugg{The role of detection resolution is illustrated in Fig.~\ref{Crit}, which shows
the probability density $p(x|a)$. The corresponding Fisher information density,
\[
F(x|a) = \frac{(\partial_x p(x|a))^2}{p(x|a)},
\]
exhibits a complementary behavior (not shown): it is enhanced in regions where
the probability density is small and suppressed where the probability density is large.
A more stringent requirement than the sampling condition arises from the need
to resolve the local structure of the probability density. Let us consider a
detector with finite resolution, such that the measurement outcomes are
collected in bins of width $\delta x$. The number of detected events in a bin
centered at $x$ is then approximately $k(x) \approx n\,p(x)\,\delta x$, with
statistical fluctuations of order $\sqrt{n\,p(x)\,\delta x}$. 
}

\sugg{ 
In order to resolve the local variation of the probability density, the change
of the expected counts across the bin must exceed these fluctuations. The
variation of the probability density over the bin is given by
$\delta p(x) \approx |\partial_x p(x)|\,\delta x$, which leads to a change in
the expected counts $\delta k \approx n\,|\partial_x p(x)|\,\delta x^2$. The
condition $\delta k \gtrsim \sqrt{n\,p(x)\,\delta x}$ therefore yields
\begin{equation}
n \gtrsim \frac{p(x)}{(\partial_x p(x))^2}\,\frac{1}{\delta x^2}
= \frac{1}{F(x|a)\,\delta x^2}.
\end{equation}
This condition must be satisfied in all regions that contribute significantly
to the Fisher information. In oscillatory distributions, these contributions
originate from regions where the probability density is small, which
implies that the required number of samples can become large. Together with the
sampling requirement $\delta x \lesssim 1/N$, this shows that resolving the
fine structure of the likelihood requires both sufficiently fine binning and a
sufficiently large data set.
This analysis highlights that Fisher information characterizes local sensitivity,
but does not account for the statistical cost of resolving the corresponding
structure. {\em In particular, regions that contribute most significantly  to the Fisher
information are the most demanding to probe experimentally.}
}
\sugg{
The  role of the data sample and noise model is therefore crucial for quantum sensing: Increasing the number of samples improves robustness but tends to recover classical scaling, while operating with fewer samples may, in principle, access regimes of enhanced sensitivity, albeit at the cost of increased susceptibility to noise and estimator instability. The optimization of this trade-off leaves room for potential quantum advantage but does not imply a genuine surpassing of the Heisenberg scaling with respect to total resources. Misinterpreting the parameter $N$ as an independent resource may therefore lead to spurious claims of super-Heisenberg resolution.
Apparent super-Heisenberg scaling in criticality-based quantum metrology has been shown to disappear once the full physical resources, such as the time required for state preparation, are taken into account~\cite{Criticality}. This observation is consistent with the general perspective adopted here since total resources scales with the time. 
}


\sugg{
An additional trade-off arises from the signal yield associated with generating
the oscillatory response. In optical systems near resonance, a rapid variation
of the refractive index (real part of the susceptibility) is typically
accompanied by strong absorption (imaginary part), leading to a reduced signal
yield. This analogy highlights that enhanced sensitivity is often accompanied by
additional constraints that limit the effective information gain and must be
incorporated into a complete inferential analysis.
}


\sugg{ 
\section{Summary of operational requirements for meaningful quantum advantage}
\label{Sum}
}

\sugg{
Based on the analysis presented in this work, we identify several necessary conditions for a meaningful and operational assessment of quantum sensing protocols, particularly in contexts where Fisher information is used as a performance benchmark:
}

\begin{enumerate}

\item  \sugg{ \textbf{Estimator-based evaluation.} 
Performance must be quantified using explicitly constructed estimators and their variance. 
Fisher-information-based bounds alone do not constitute an operational measure of achievable precision.
}

\item \sugg{ \textbf{Finite-sample consistency.} 
The number of samples required to reach the asymptotic regime must be taken into account, as the Fisher information per detection event does not determine a reliable measure of performance.
}

\item \sugg{ \textbf{Consistent resource accounting.} 
All resources must be incorporated into a unified framework, including both the cost per experimental run and the total number of repetitions.
}

\sugg{
\item \textbf{Global identifiability.} 
Ambiguities arising from multi-modal likelihood functions require additional resources to resolve and must be explicitly accounted for.
}

\sugg{
\item \textbf{Criticality.} 
Criticality-based protocols rely on detecting rapidly varying parameters associated with low-probability events; their performance is therefore highly sensitive to noise and finite resolution, and does not necessarily imply operational quantum advantage.
}

\sugg{
\item \textbf{Signal yield and efficiency.} 
State preparation, coupling, and detection efficiencies must be included in the overall performance assessment, as they directly affect the achievable precision.
}

\sugg{
\item \textbf{Role of Fisher information.} 
Fisher information per detection should be regarded as a tool for the design and optimization of sensing protocols, rather than as a standalone performance metric. Any performance claim must be supported by an explicitly constructed estimator.
}
\end{enumerate}

\sugg{
Only when all these aspects are consistently incorporated can claims of quantum-enhanced performance be considered operationally meaningful. Neglecting any of these aspects may lead to overly optimistic or non-operational conclusions about quantum advantage.
}

\section{ Conclusion} \label{Con}

Quantum technologies draw on the combined intellectual heritage of quantum mechanics and mathematical statistics, forming a natural meeting ground for physicists and statisticians. In this work, we have developed a unified inferential framework for quantum sensing that explicitly incorporates finite data, estimator bias, prior information, and operational resource constraints stressing the fact that a natural unit for estimation  is data set needed for construction of estimator, not a single detection itself.

Because our discussion begins with Fisher’s foundational  work, it is fitting to return to his own remark on the application of statistical methods to quantum theory, given on page 315 of Ref.~\cite{Fisher_22}:
“... Planck’s development theory of quanta ... in all these applications the methods employed have been, from the statistical point of view, relatively simple.”  A century later, this assessment remains strikingly accurate. In large parts of modern quantum metrology, simplified conclusions are still drawn from information-theoretic bounds alone, and exotic quantum states are often claimed to provide dramatic advantages over classical, Gaussian resources. NOON states have become emblematic of this narrative. Yet, in the vast majority of studies, performance is assessed exclusively through QFI, without constructing explicit estimators, quantifying the required resources, or incorporating prior information and  appropriate benchmarks. 

Our results show that neglecting these elements can qualitatively alter the conclusions. Once priors are included and resources are accounted for in a consistent manner, highly non-Gaussian states may offer  no practical advantage over standard strategies. Without an explicit end-to-end inferential analysis, claims of quantum advantage remain statements about mathematical bounds rather than about achievable metrological performance. Clarifying these long-standing misconceptions and redirecting attention from idealized figures of merit to operationally attainable sensing capabilities is essential if the genuine potential of quantum sensing is to be reliably identified and ultimately realized.

\section*{Acknowledgment}
We acknowledge the support of the  Czech Science Foundation   under the grant agreement 26-22242J.

\section*{Author Contribution}
Z.H. conceived and developed the theoretical framework. J.R. performed the numerical analysis. Both authors contributed to manuscript writing and editing.

\section*{Data Availability}

This work is theoretical and no experimental datasets were generated or analyzed. Data supporting the numerical analysis are available from the corresponding author upon reasonable request.

\bibliography{bibOlomouc}

\end{document}